**Classifying active and inactive states of growing rabbits from accelerometer data using machine learning algorithms**

Mónica Mora[*], Lucile Riaboff [#,†,§], Ingrid David[#], Juan Pablo Sánchez[*] and Miriam Piles[*]

[*] Animal Breeding and Genetics, Institute of Agrifood Research and Technology (IRTA), Caldes de Montbui, 08140 Barcelona, Spain.

[#] GenPhySE, Université de Toulouse, INRAE, ENVT, 31326 Castanet Tolosan, France

[†] School of Computer Science, University College Dublin, Dublin, Ireland

[§] VistaMilk SFI Research Centre, Ireland

Corresponding author: monica.mora@irta.cat

**Funding**

This study was part of the project PID2021-128173OR-C21 (GENEF3) and MM is a recipient of a "Formacion de Personal Investigador (FPI)" associated with the research project RTI2018-097610R-I00.

**Lay Summary**

This study explores how wearable accelerometers, small devices that measure acceleration, can help monitor the activity of growing rabbits. We equipped 16 rabbits with these devices and filmed them for two weeks. By watching the videos and using a special software we figure out what the rabbits were doing – things like lying down, eating, moving around, and more. These activitties were grouped into two states: active or inactive. Then, this information along acceleration data was used to teach a computer program to recognize when the rabbits were active or not. This technology offers a reliable way to understand rabbit behavior, which could lead to better management practices in animal production.

## Teaser text

This research showcases the effectiveness of wearable accelerometers combined with machine learning in accurately monitoring rabbit inactivity levels.

## Abstract

Using wearable accelerometers is gaining traction in research and animal production management for monitoring animal behaviour. In this study, the objective was to automatically detect rabbit activity/inactivity states from accelerometer data in growing rabbits. For that purpose, 16 animals were equipped with an accelerometer and filmed for 2 weeks. A total of 10 hours of video across all the rabbits were annotated manually using the Boris software, identifying 6 classes of different behaviours: lying, eating, moving, grooming, walking and drinking which were grouped into two classes: active and inactive. Accelerometer signal and video annotations were manually synchronized. The static and dynamic components of the signal were isolated by applying a low-pass and high-pass filter and 4 additional time series were derived from these components. The signal was segmented into time windows of different sizes: 1, 3, 5, 7 and 9 seconds. For each window, a total of 41 features were extracted in the time and frequency domain. Different subsets of data containing an increasing number (from 5 to 25 in steps of 5) of the most informative features identified with random forest (RF) were used to train a binary classification model (inactive as a positive class). The classification performance of RF, support vector machine (SVM) and gradient boosting (GB) was evaluated. A nested cross-validation (CV) with an outer Leave-One-Animal Out CV and an inner threefold CV for hyperparameter tunning was implemented. The same resampling was implemented for each window size and each classifier so that the models were evaluated with the same data sets. The performance was evaluated on the test datasets using different metrics: precision, recall, F1 score and accuracy. Results showed that the classifiers perform very similarly. With the best configuration (window size of 9 s and with the 5 most important features) the RF model reaches a median precision of 1 (Q1=0.99, Q3=1) and a median recall of 0.93 (Q1=0.89, Q3=0.97). These results showed that the model is highly reliable in correctly classifying positive instances. Additionally, achieving a recall of 0.93 emphasizes the model's effectiveness in capturing a substantial portion of positive instances. Accelerometers combined with machine learning models therefore hold great promise for monitoring rabbit activity and for a range of applications in animal science and behaviours.

**Keywords:** accelerometer, rabbit, classification, monitoring, machine learning

## List of abbreviations

CV: cross-validation

FN: false negative

FP: false positive

FPR: false positive rate

GB: gradient boosting

LOAOCV: leave-one-animal out cross-validation

PLF: precision livestock farming

Q: quartile

RF: random forest

ROC: receiver operating characteristics

SMOTE: synthetic minority over-sampling technique

SVM: support vector machine

TN: true negative

TP: true positive

TPR: true positive rate

**Introduction**

Monitoring livestock behaviour is crucial for promptly detecting epidemic outbreaks and/or technical and environmental issues, such as drinking trough failures and inadequate temperatures. Furthermore, the activity level has been shown to be related to feed efficiency in beef (Haskell et al., 2019), making this information valuable for inclusion in breeding programs to enhance this trait, which stands as one of the primary objectives for the majority of livestock species.

In the case of rabbits, it would be essential to monitor their behaviour. Understanding how their activity patterns relate to their health and feed efficiency could provide valuable insights for improving their management and welfare. However, continuous monitoring of the animals by humans is time-consuming and labour-intensive, making it impossible to monitor them 24/7. Therefore, the integration of new and advanced technologies in livestock is getting highly relevant for ensuring sustainable and efficient production systems. Precision livestock farming (PLF) aims to create smart farms that use these technologies to continuously and automatically monitor the animals (Berckmans, 2017). In the last few years, different types of sensors and devices have been proposed to achieve this goal. Among these tools, wearable sensors containing accelerometers can be employed to gather information about the activity at the individual level (Chapa et al., 2020). An accelerometer is an electromechanical device that measures acceleration forces, which is the rate of change of velocity of an object. Moving the sensor generates a voltage by applying stress to the small crystals it contains. In a tri-axial accelerometer, acceleration is measured along the three-dimensional axis (x, y and z) (Benjamin and Yik, 2019). The data recorded by the accelerometer provides information on the animal's movements, including how often and how vigorous they are, which in turn can provide insights into animal behaviour, health conditions, pain or discomfort. Accelerometers have gained popularity due to their compact size and affordable cost, being utilised in numerous livestock applications. Above all, they have found significant applications in large animals and/or those that are outdoors (Martiskainen et al., 2009; Shahriar et al., 2016). For example, accelerometers have been employed in sows that are housed individually in gestating stalls (Ringgenberg et al., 2010). In cattle, behaviours such as rumination, feeding time and grazing time have been the most studied based on accelerometer data (Chapa et al., 2020). However, to the best of our knowledge, wearable accelerometers have not been used yet to monitor the behaviours of rabbits. Accelerometers could then be utilized to get information on activity patterns throughout the day in their housing system and to assess the modification in behaviour related to changes in their environment, animal growth, etc. However, getting behaviours from raw accelerometer signals requires applying appropriate experimental design and analytical methods (Riaboff et al., 2022). Indeed, different dynamical properties and patterns corresponding to an observed behaviour can be identified in the raw accelerometer signal. Supervised machine learning algorithms demonstrated effectiveness in classifying accelerometer signals into pre-defined behaviour categories (Vázquez Diosdado et al., 2015; Riaboff et al., 2020).

To train and validate such algorithms, accelerometer signals and their corresponding observed behaviours are required. Animal behaviours can be recorded through direct observation (Rayas-Amor et al., 2017; Rodriguez-Baena et al., 2020) or with images captured by a camera (Cornou and Lundbye-Christensen, 2010; Thompson et al., 2019). Then, the signal is segmented into time windows. For each window, a set of statistical features is calculated. The classifier is trained using the combination of these statistical features as predictor variables and the corresponding behavioural annotations as the target. Once the model is trained, it can be applied to new accelerometer signals from new individuals to classify and monitor their behaviours. In this way, attaching an accelerometer to the rabbits would allow for obtaining individual patterns of activity, enabling their monitoring over long periods.

The present study aimed to develop and validate a model to classify active or inactive states over 15 days using accelerometer data from growing rabbits.

## Material and methods

### Ethics approval

The experimental procedures used in the research complied with the European directive 2010/63/EU and the Spanish guidelines for the care and use of animals in research (B.O.E. number 34, Real Decreto 53/2013). Animals were submitted to an experimental procedure approved by the Ethical Committee of IRTA and the Regional Government of Catalonia.

### Accelerometers

The accelerometers used in this study were the AX3 dataloggers (https://axivity.com/product/ax3). They are MEMS 3-axis accelerometers with onboard memory (512 MB), measuring 23 × 32.5 × 8.9 mm and weighing 11 g. The accelerometer has configurable sample rates (frames per second) and sensitivity (+/- 2 g, +/- 4 g, +/- 8 g and +/- 16 g; 1 g =~9.81ms$^{-2}$) so that both can be adapted depending on the targeted activities. In the present study, the sampling rate was set up to 25 Hz and the sensitivity to ± 8 g. With this configuration, the expected battery life was 30 days. Configuration and data download were performed by connecting each AX3 to a computer using a USB port and the OmGui software. The data of each accelerometer was downloaded at the end of the follow-up period for each animal.

**Methodology**

In this section, the three main steps of the methodology are described: data collection, data preprocessing and model fitting and classification (Figure 1).

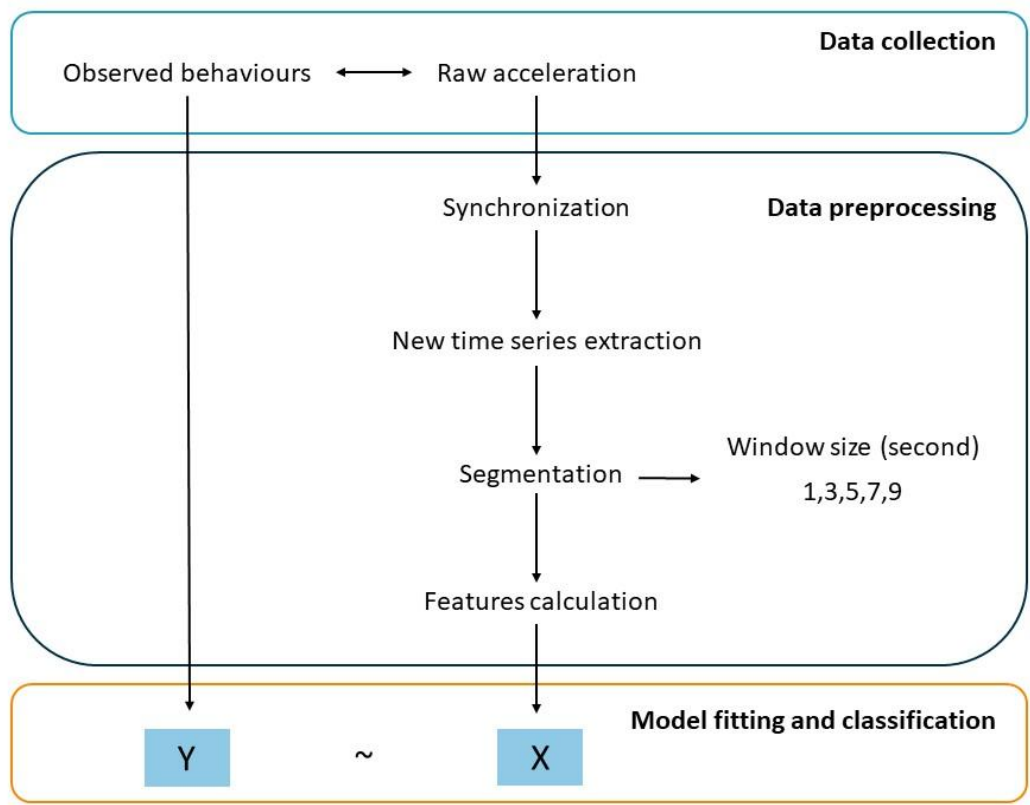

Figure 1. Overview data processing

*Step 1: data collection*

The experiment was carried out at IRTA's nucleus farm at Caldes de Montbui, Spain. In this farm, the photoperiod was set to 16 light h/d (from 07:00 to 23:00h) and the environmental temperature was kept at 20 ± 2 °C. The animals included in this study come from a sire line commonly used in rabbit farms. The animals were housed in groups of 6 kits in cages measuring 76 cm width x 100 cm length x 32 cm height. Each cage was equipped with an electronic feeder that allowed one rabbit to eat at a time ensuring accurate identification of the animal. The study considered a total of 16 animals belonging to 2 batches and born between November 2022 and February 2023. The observation period was from 37 to 51 days of age.

In each batch, one animal from each cage (8 cages in total) was randomly selected to be equipped with the accelerometer for 15 consecutive days. The tri-axial accelerometer was enclosed into a resistant plastic box and fixed to the rabbit's back using an elastic band (Figure 2). All the sensors were attached in the same way for all the animals, i.e., with the same orientation. This ensured that the x-axis detected the backward-forward direction, the y-axis detected the left-right direction and the z-axis detected the down-up direction whatever the rabbit.

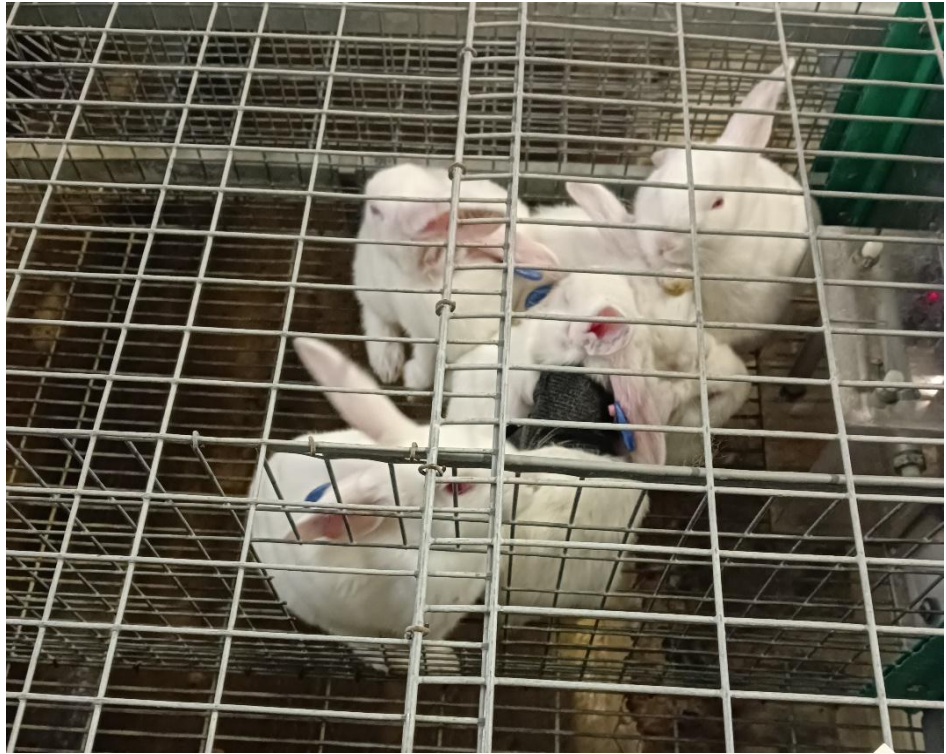

Figure 2. Rabbit equipped with the accelerometer.

The rabbits were videotaped throughout the two-week experiment in each of the batches using a 2D camera placed on the top of the cages. Videos of rabbits equipped with the accelerometer were observed for at least 35 minutes and a maximum of 39 minutes (not entirely sequential). All the pre-defined behaviours that took place during that time were manually annotated by the same person according to an ethogram (Table 1) and the BORIS software (Friard and Gamba, 2016). In the end, animals' behaviours on various days of the experiment and at different moments of the day were annotated, resulting in a total of ~10 hours of video annotations across the 16 rabbits.

Table 1. Definition of the different behaviours annotated.

| Behaviour | Classification description | State | Hours of observation |
| --- | --- | --- | --- |
| Lying | The rabbit is lying down | inactive | 5.34 |
| Eating | The rabbit is inside the electronic feeder | inactive | 1.75 |
| Moving | The rabbit is moving without changing its position | active | 1.68 |
| Grooming | The rabbit is licking itself | active | 0.67 |
| Walking | The rabbit is moving while changing its position | active | 0.14 |
| Drinking | The rabbit is drinking | inactive | 0.34 |

Due to the lack of representation of some predefined behaviours, they were grouped according to the state of the animal as active or inactive. Thus, it was considered that the animal was active when it was moving, walking and grooming, and inactive when it was lying, eating and drinking. The last two behaviours were included in the inactive class because of the little movement of the animal's back that produced a flat signal in the accelerometer.

*Step 2: data preprocessing*

In this section, the different phases of data preprocessing after data collection are described. Four sequential operations were carried out: i) synchronization of the accelerometer signal with the behaviour annotations; ii) new time series extraction; iii) accelerometer signal segmentation into time windows; and iv) calculation of statistical features in each of the time windows. All the codes were run with the programming language Python v3.10.11. The main packages used were SciPy (Jones et al., 2001) and Scikit-learn (Fabian Pedregosa Gaël Varoquaux, 2011).

*Synchronization.* The time stamps of accelerometer signals and behaviour annotations were synchronized to guarantee that the annotations corresponded to the correct sequences of the accelerometer signal. The synchronization was possible thanks to a pattern in the signal created by shaking the accelerometer in front of the cameras before and after the experiment. Thus, in the videos, accelerometer shaking could be observed, allowing for accelerometer timestamp synchronization with the videos. As a result of this process, a dataset was generated in which behaviour annotations for each rabbit matched the corresponding accelerometer data sequences. However, the accelerometer time drift (±1.7 seconds per day maximum; AX3 user guidebook: https://axivity.com/userguides/ax3/), i.e., the fact that the accelerometer clock is not able to maintain constant precision over time, moving forward or falling behind, led to a time difference between the accelerometer time-series and the behaviour annotations, even after the synchronization process). Therefore, each continuous sequence of time series associated with annotations (~ 35-39 minutes) was visually inspected. Any temporal mismatch identified between the annotations and the time series was noted. The alignment was then corrected for each sequence to ensure that it was perfectly synchronised with the associated behaviour.

*New time series extraction.* From the measurements in each of the three axes (x,y and z), a fourth orientation-independent one was calculated as follows:

$$a_{mag} = \sqrt{a_x^2 + a_y^2 + a_z^2}$$

Where $a_x$, $a_y$ and $a_z$ are the acceleration on the x-axis, y-axis and z-axis, respectively. This variable is called "the magnitude".

On the other hand, the acceleration measured by the accelerometer in the three axes may be separated into both static and dynamic acceleration. The static acceleration registers the orientation of the animal, while the dynamic gets information about the energy expended without the 'bias' of the gravitational force. To separate static from dynamic acceleration, filters are typically applied to the raw acceleration. In the present study, low and high-pass filters were used to get the static and the dynamic acceleration, respectively. Considering that the static component is constant and thus leads to frequencies ~ 0, usually, the cut-off frequency used is 0.3 Hz (Smith et al., 2016; Riaboff et al., 2020). From the static acceleration, the angles around x (**roll)** and y (**pitch**) were calculated to estimate the accelerometer orientation following the definition presented by Rautiainen et al. (2022). From the dynamic acceleration, the norm of the dynamic acceleration (**VeDBA**) and the sum of the absolute value of dynamic acceleration (**OBDA**) were computed. The formulas for all these features are as follows:

$$\text{pitch} = -\text{atan}\left(\frac{\text{static}_x}{\sqrt{\text{static}_y^2 + \text{static}_z^2}}\right)$$

$$\text{roll} = \text{atan2}(\text{static}_y, \text{static}_z)$$

$$\text{VeDBA} = \sqrt{\text{dynamic}_x^2 + \text{dynamic}_y^2 + \text{dynamic}_z^2}$$

$$\text{OBDA} = |\text{dynamic}_x| + |\text{dynamic}_y| + |\text{dynamic}_z|$$

*Time-windows creation.* The accelerometer signal was split into time windows of the same size. As the movement of rabbits is brief, window sizes smaller than 10 seconds were created (Riaboff et al., 2022). To determine the optimal window size, various short window sizes were tested: 1, 3, 5, 7 and 9 seconds. Windows with a mix of both active and inactive classes were removed from the analysis. Thus, each window was associated with just one of the classes. In Table 2, for each window size, the total number of obtained windows is shown together with the number of them belonging to the inactive class.

Table 2. Total number of windows and percentage of windows in the inactive class.

| Window size | Total windows | Percentage windows inactive |
|---|---|---|
| 1 s | 24,685 | 0.77 |
| 3 s | 7,471 | 0.80 |
| 5 s | 4,052 | 0.82 |
| 7 s | 2,676 | 0.86 |
| 9 s | 1,941 | 0.86 |

*Features calculation.* For each time window, features were computed on each of the time series in the time and frequency domain. A total of 41 features were computed. Table 3 shows all the features obtained and their formulas. In the time domain, the features calculated were related to position parameters like the mean and median, as well as dispersion parameters like the standard deviation, the minimum, the maximum and the range. Kurtosis and skewness were also computed in the time domain. Skewness measures the lack of symmetry of the data distribution while kurtosis measures whether the data are heavy-tailed or light-tailed relative to a normal distribution. Another feature typically used is the motion variation which indicates the rate of change in the signal: a more energetic movement will lead to higher motion variation. Finally, in the frequency domain, the spectral entropy was calculated only for the magnitude axis. It refers to a measure of the random process uncertainty in the accelerometer signal frequency distribution. Only the range was computed for the pitch, roll, OBDA and VeDBA time series.

Table 3. Formulas of the extracted features from the acceleration data.

| Feature | Formula |
|---|---|
| Average | $\bar{A}_a = \dfrac{1}{M}\sum_{i=1}^{M} a_i$ |
| Median | The median of the axis within the window |
| Standard deviation | $\sigma_a = \sqrt{\dfrac{\sum_{i=1}^{M}(a_i - \bar{A})^2}{M}}$ |
| Minimum | The minimum of the axis within the window |
| Maximum | The maximum of the axis within the window |
| Range | $\text{Range}_a = \text{Maximum}_a - \text{Minimum}_a$ |
| Kurtosis | $\beta_{1,a} = \dfrac{1}{M}\sum_{i=1}^{M}\left(\dfrac{a_i - \bar{A}_a}{\sigma_a}\right)^3$ |
| Skewness | $\beta_{2,a} = \dfrac{1}{M}\sum_{i=1}^{M}\left(\dfrac{a_i - \bar{A}_a}{\sigma_a}\right)^4$ |
| Motion variation | $MV = \dfrac{1}{M}\left(\sum_{i=1}^{M-1}\|a_{x,i+1} - a_{x,i}\| + \sum_{i=1}^{M-1}\|a_{y,i+1} - a_{y,i}\| + \sum_{i=1}^{M-1}\|a_{z,i+1} - a_{z,i}\| + \right)$ |
| Spectral entropy | $H_s = -\sum_{k=1}^{N} P_k \ln(P_k)$  $P_k = \dfrac{\|\lambda_k\|^2}{\sum_i \|\lambda_i\|^2}$ and for the frequency $\lambda_k$ where N is the number of points in the signal spectrum |

Note: The formulas are depicted as a function of the acceleration along any axis denoted as a, while M is the total number of samples in the window.

*Step 3: Model fitting and classification of the animal state from the accelerometer signal*

The goal of this step was to train a machine learning model using the features calculated in the previous step to classify the animal state (active or inactive) during every time window. The inactive class was considered as the positive class.

Three machine learning algorithms were tested and their classification performance was evaluated: Random Forest (**RF**) (Breiman, 2001), Support Vector Machine (**SVM**) (Cristianini and Ricci, 2008) and Gradient Boosting (**GB**) (Friedman, 2000). RF is an ensemble learning method combining several classification trees. Each decision tree is fitted to a set of subsamples randomly selected through a bootstrapping procedure and using randomly selected subsets of the predictor variables. The global prediction is then obtained as the majority vote across all the predictions from all trees. GB builds a series of weak decision trees, where each subsequent tree corrects the errors made by the previous ones. The final prediction is a combination of all the weak trees. On the other hand, SVM is a supervised learning algorithm, that aims to find the hyperplane that best separates different classes in the feature space. More information about these classifiers can be found in the literature cited above.

For each classifier, a nested resampling (Figure 3) was implemented to obtain reliable performance estimates. This procedure consists of two nested resampling loops. In the outer resampling loop, a Leave-One-Animal Out Cross-Validation (LOAOCV) was carried out. In LOAOCV, the dataset is divided into n subsets, where n is the number of animals. For each iteration of the outer loop, data points corresponding to one animal are held out as the testing dataset, and the model is trained on the remaining data points, ensuring that all the data points are used for prediction. The models were thus tested with different animals than those used for model training, preventing overfitting and over-optimistic results. This approach allowed us to evaluate the predictive power of the model while considering the variability inherent in individual animals.

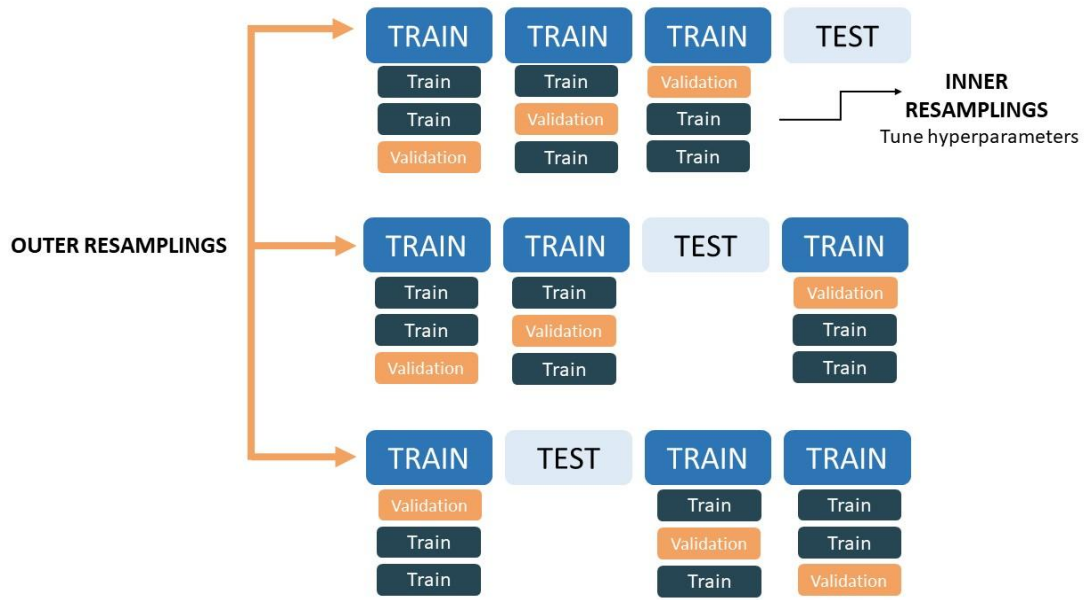

Figure 3. Nested k-fold cross-validation diagram.

In each outer training set, feature selection was carried out by selecting subsets of the most informative features using RF. The best features (i.e. the ones that contribute the most to the classification of the animal's state corresponding to every window period) were selected based on their Gini importance scores, which quantify the reduction in impurity achieved by each feature across all the decision trees. For each subset of an increasing range of the most important features (from 5 to 25, in steps of 5), a 3-fold CV was implemented in each outer training set, where one fold was used as a validation set and the remaining folds were used as an inner training set. The evaluation of this inner CV is used in a grid-search procedure to find the optimum hyperparameters of the machine learning classifiers. Each model is built and evaluated with every combination (discrete grid) of the manually specified subset of values for each hyperparameter. The discrete grid applied for each classifier is presented in Table 4. We refer to the scikit-learn library for details about hyperparameters and values (Fabian Pedregosa Gaël

Varoquaux, 2011). Once, the optimal hyperparameters were found, the model was fitted to all training data. The output of the model is the probability that the time window belongs to a given class. The optimal threshold for which a window is classified as inactive was calculated from the Receiver Operating Characteristics (**ROC**) curve directly. This curve represents the true positive rate (TPR) or recall *versus* the false positive rate (FPR) at each threshold setting.

$$\text{TPR or recall} = \frac{\text{TP}}{\text{TP} + \text{FN}}$$

$$\text{FPR} = \frac{\text{FP}}{\text{FP} + \text{TN}}$$

The geometric mean or G-mean is a metric for imbalanced classification that, if optimized, will seek a balance between the TPR and the specificity, which is 1-FPR. Then, when the ROC curve is calculated, the G-mean can be calculated for each threshold:

$$\text{G} - \text{mean} = \sqrt{\text{TPR} * (1 - \text{FPR})}$$

The optimal threshold is determined by the probability corresponding to the largest G-mean.

The same scheme and the same data split were used to compare prediction performance in the same conditions for the different window sizes and the three classifiers.

Table 4. Hyperparameters tunned for each classifier

| Classifier | Hyperparameter | Discrete grid |
|---|---|---|
| Random Forest | **Number of estimators**: number of trees in the forest | 100, 250, 500, 800 |
| | **Max depth**: maximum depth of the tree | 2, 4 |
| | **Min samples split**: the minimum number of samples required to split an internal node | 2, 4 |
| | **Min samples leaf**: the minimum number of samples required to be a at leaf node | 4, 8 |
| Gradient Boosting | **Number of estimators**: the number of trees to perform | 100, 250, 500, 800 |
| | **Learning rate**: shrinks the contribution of each tree | 0.01, 0.1 |
| | **Max depth**: maximum depth of the individual trees | 2, 4 |
| Support Vector Machine | **C**: regularization parameter | 1, 3 |
| | **Kernel** | Linear, RBF |
| | **Gamma**: kernel coefficient | 0.1, 0.4 |

**Evaluation of the models**

The performance was evaluated in each of the test datasets using different metrics: recall, precision, F1 score and accuracy. The confusion matrix was also evaluated. Recall measures how often the models correctly identify inactive windows (true positive) from all the actual positive windows in the dataset. Precision measures how often the model correctly predicts inactive windows. It is defined as follows:

$$\text{Precision} = \frac{\text{TP}}{\text{TP} + \text{FP}}$$

The F1 score was computed to assess the overall performance of a classification model. This metric provides a single value that balances the recall and the precision:

$$F1\ Score = \frac{2 \times Precision \times Recall}{Precision + Recall}$$

The accuracy is the proportion of correct predictions made by the model out of the total predictions made:

$$Accuracy = \frac{TP+TN}{TP+FN+TN+FP}.$$

## Results

### Model performance

In this section, the performance of each classifier for each window size and each feature subset size is evaluated. It is important to note that a LOAOCV procedure was conducted in a combination of the learner, the window size and the feature subset size. Therefore, there were 16 prediction performances (one for each animal) for each combination.

For the same window size, similar results were obtained regardless of the number of most important features (see figures in supplementary material, Figure S1, S2, S3, S4, and S5). Therefore, here, we present the performance of the simplest and fastest model to train, which corresponds to the model trained using the top 5 features according to RF. The model performance of each classifier for each window size with this subset of the most important features is presented in this section.

The distribution of precision, recall and F1 score with the top 5 features depending on the window size for each machine learning model is presented in Figure 4. The median of the precision, recall and F1 score increased as the window size increased, regardless of the machine learning model. However, the differences were not substantial.

Similarly, no differences were observed among the three classifiers for the same window size. From now on, we will focus on the results obtained for each machine learning model with a window size of 9 seconds and the 5 top RF features. In that configuration, a median F1 score of 0.96, ranging from 0.88 to 0.99, was obtained with RF. The median (Q1, Q3) of F1 scores for the classifiers SVM and GB were 0.95 (0.90, 0.98) and 0.95 (0.93, 0.97), respectively. Regarding the precision, the median (Q1, Q3) was 1 (0.99, 1) for RF, SVM and GB. This means that 100% of the time when a window was predicted as the positive class (inactive), it truly belonged to this class. In addition, the median (Q1, Q3) recall for RF was 0.93 (0.89, 0.97). That means that 93% of the windows corresponding to inactive periods were detected by the model. The recall median (Q1, Q3) for SVM and GB was 0.91 (0.84, 0.97) and 0.92 (0.87, 0.95), respectively.

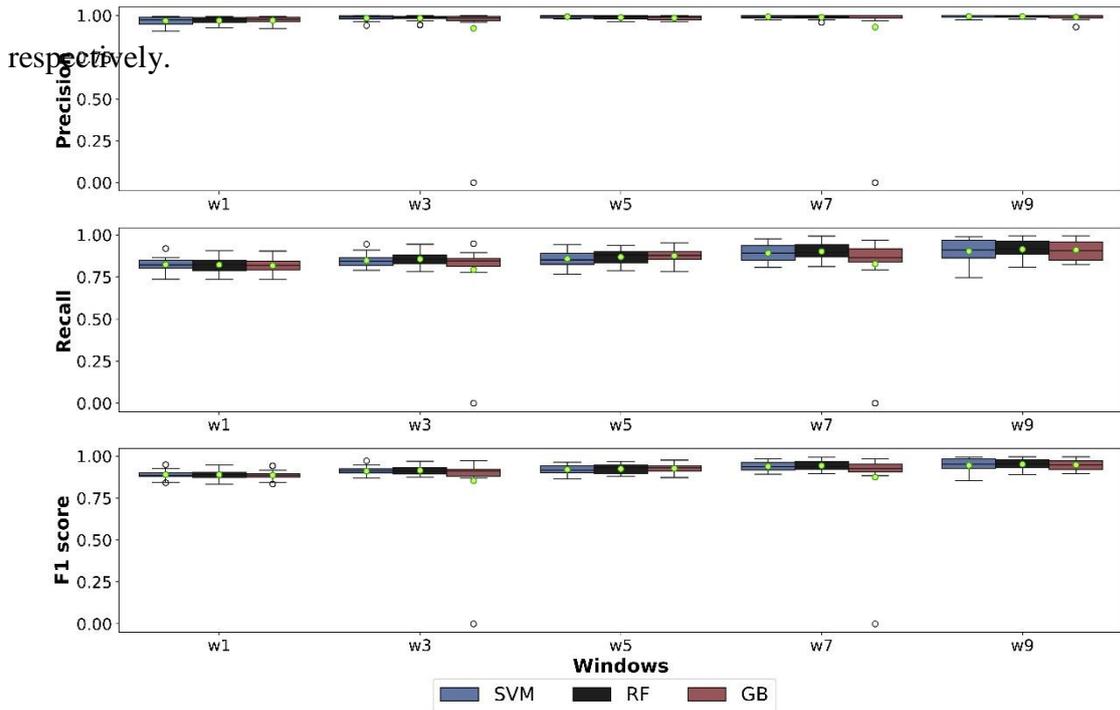

Figure 4. Distribution of precision, recall and F1 score for each model for each window size with the top 5 features. The green dot represents the mean across the 16 outer loops.

Regarding the accuracy of the model, with a window size of 9 seconds and the 5 top best features the median accuracy (Q1, Q3) for RF was 0.93 (0.89, 0.96), similar to the other classifiers (results not shown).

Since there is no substantial difference between the classifiers, only the confusion matrix for RF is presented (Figure 5), showing the median values, as well as the Q1 and Q3 of the 16 folds. Regarding the positive class, i.e., inactive, 92 windows were classified as positive, all of them TP. Only, 8 out of 100 inactive windows were misclassified as active windows (FN). Focusing on the negative class, out of the total of the 19 windows classified as active 42% of them belonged to the inactive class. If the active state were considered as the positive class, the median recall would be 100% $\left(\frac{11}{11+0}\right)$ and the median precision of 58% $\left(\frac{11}{11+8}\right)$. The most important features were the motion variation across the x, y, z and magnitude axes.

|  | | Predicted | |
|---|---|---|---|
|  | | active | inactive |
| **Actual** | active | TN<br>**11** (5, 16.75) | FP<br>**0** (0, 1) |
|  | inactive | FN<br>**8** (3.5, 11) | TP<br>**92** (80, 105) |

Figure 5. Confusion matrix for random forest at a window size of 9 seconds with the top 5 more important features. The median values (Q1, Q3) of the 16 folds are depicted.

**Inference of active/inactive states**

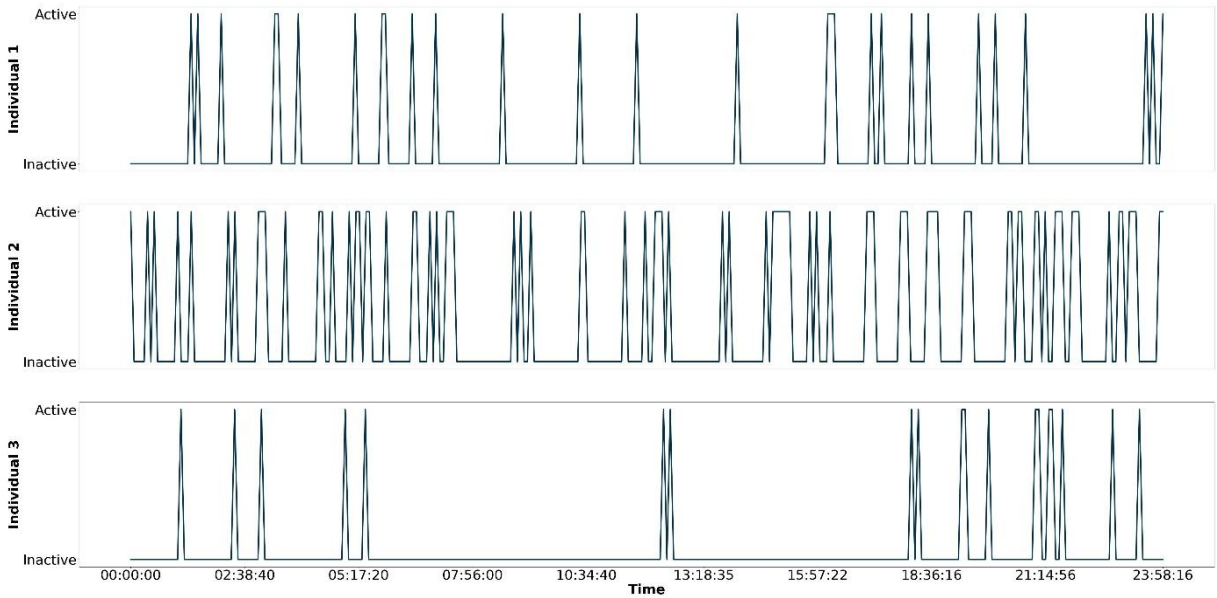

Figure 6. Daily pattern activity of three individuals.

The best RF model based on the F1 score was used to perform inference on the activity level of three randomly selected rabbits throughout a day. The model's input consisted of raw accelerometer signals segmented into equal-sized windows of 9 seconds each. The daily pattern activity of these rabbits is presented in Figure 6. In this figure, it can be observed that the most significant activity patterns occurred during the night. This is particularly evident in individuals 1 and 3, where intense activity concludes around 5/6h in the morning and resumes around 18h.

## Discussion

Our analysis aimed to train a model capable of classifying two general states of growing rabbits, active and inactive, based on accelerometer data. To find the best possible classification performance, multiple machine learning algorithms have been tested, with their respective hyperparameter tuning, on top of a variety of window sizes for the accelerometer signals.

Firstly, our study showed that accelerometers attached to the back of rabbits in cages can be used to reliably classify inactive states. With a window size of 9 s and the top 5 features, the RF model reaches a median precision of 100% and a median recall of 93%. In addition, the models were validated in a configuration where the rabbits used for model training have not been used for model testing, suggesting that the model also has a good genericity (Riaboff et al., 2022). Furthermore, it is noteworthy to point out that the rabbits did not struggle to wear the sensor. It did not affect their behaviour or that of their cage mates, no fights or attempts to remove the device were appreciated. At the end of the experiment, the accelerometers were retrieved in perfect condition, with no hardware damage, making it a viable option to measure the activity of growing rabbits. To the best of our knowledge, this is the first attempt where wearable accelerometers have been used to monitor the behaviours of growing rabbits, thus highlighting the potential of such technologies in that field.

Even if the models' performance are really good, several improvements could be proposed. First, it was not possible to simultaneously test different accelerometer configurations due to the limited number of sensors at our disposal. Based on results obtained from other species and following the device instructions, the sampling rate was set up to 25 Hz and the sensitivity to ±8g. However, for future analyses, it would be advisable to test multiple configurations and assess which is the best one for the case study.

As for the placement, in rabbits, the most feasible option was to attach it to their backs, at the level of the rib cage, using an elastic band Figure 2. This arrangement enables the sensor to move with the general movement of the rabbit but may fail to catch activities such as eating when there is only head movement. Thus, the signal between the animal lying down and eating can be hardly distinguished. In the case of the drinking activity, since the position of the nipple is on top of the cage, the rabbit had to stand up, which should be reflected on the accelerometer data. The decision of where to place the sensor is therefore of high relevance as it will determine which behaviours will be measured with higher precision (Chapa et al., 2020). In the literature, some work can be found where more than one accelerometer was employed, demonstrating an improvement in predicting behaviours *versus* the only one sensor approach (Ringgenberg et al., 2010; Thompson et al., 2019). However, in growing rabbits, this option seems difficult to achieve.

In the process of synchronization, an average gap between the camera clock and accelerometer clock was observed in each continuous sequence due to the time drift. Thus, the impact that it might have on the correspondence between the accelerometer signal and the annotated behaviours can be significant if not corrected. In this study, each continuous sequence was manually corrected to ensure a perfect match between the signal and the annotations but this is time-consuming and labour-intensive, which may make this type of study difficult to achieve. One solution to this problem of accelerometer clock drift is to equip the animals and film them for several hours (~8 hours), then reset the cameras and sensors, and repeat the test the next day, since drift accumulates over time.

In the context of model training, the number of samples is also highly important. It depends on the amount of data available for each behaviour but also of the chosen window size; bigger window sizes result in smaller sample sizes and *vice versa*. Furthermore, for large window sizes, there is a greater likelihood of having several behaviours in the window and therefore, of excluding it from the dataset for the analysis. If these windows are not removed but instead the most likely class is assigned, the signal corresponding to different behaviours would be present in the same window, leading to noise. With a window size equal to 9 seconds, 1,941 windows were generated, 86% of them belonging to the inactive class, resulting in a class imbalance problem. For that purpose, techniques for over-sampling and under-sampling could be applied (Turner et al., 2022). In the present study, multiple Synthetic Minority Over-sampling Technique (SMOTE) variants were tested (results not shown), nevertheless, they did not enhance the performance of the overall model. Results could be improved with additional data annotations and with a better representation of all the classes.

Multiple features were extracted from the segmented accelerometer signal, as typically done in the field to predict a variety of behaviours from accelerometer data (Shahriar et al., 2016; Lush et al., 2018; Pandey et al., 2021). A parallel approach can be to apply neural networks, as done by Turner et al. (2022), to the accelerometer time series, thus avoiding knowledge-based feature extraction. They reported that neural networks outperformed classical machine learning algorithms. However, this kind of approach requires a substantial dataset size.

In the present study, the model accuracy was greater than that of the majority class proportion. Regarding the trade-off between precision and recall, with RF and a window size of 9 seconds, we can predict inactive behaviours with a median precision of 1 (Q1=0.99, Q3=1) and a median recall of 0.93 (Q1=0.89, Q3=0.97). As already explained, precision refers to the proportion of TP among all positive predictions and recall refers to the proportion of TP among all the actual positive cases. In this case, the model is highly reliable in correctly classifying positive instances and it is capable of detecting 93% of the positive instances. The main limitation to achieving higher recall values is that the signal in some periods classified as inactive is similar to that obtained in some periods classified as active. This may be a consequence of multiple factors, such as having mixed data in the inactive class (the sensor might move when the animal is drinking or eating), the placement and strapping of the sensor or possible light interactions between the rabbits (the sensor is slightly hit by another rabbit). On the other hand, if the active class were considered as the positive class, we would obtain a perfect recall but a median precision of 58% due to confusion between inactive instances and active ones.

A potential use of the activity patterns obtained from accelerometers in growing rabbits would be the detection of sick animals (González et al., 2008; Dittrich et al., 2019). These daily patterns, as shown in Figure 6, could be useful for implementing alarm systems for intervention if a behaviour change is observed. Moreover, understanding daily activity patterns can also assist in optimizing management strategies and identifying potential stressors or environmental factors affecting animal behaviour. On the other hand, the activity pattern has been found to be related to feed efficiency in beef steers (Haskell et al., 2019). This opens the possibility of defining new or additional phenotypes that could be included in a breeding program to select for feed efficiency.

Nevertheless, some drawbacks need to be considered regarding the use of this kind of device. The accelerometer used in this study is required to be connected to the computer to download the data. That means, that the device must be removed from the animals every time this action is needed. In rabbits, this is not a significant problem because they are small animals and easy to handle. Similarly, the anchoring system was an elastic band, which is easy to put on and take off but it involves handling the animal in a way that can be stressful for it. On top of that, a long-lasting battery and Bluetooth connection that ensures storage of data on an external database are important characteristics to be included in the device without compromising its price, size and weight.

The greatest advantage of these devices compared to other systems is that they allow the individualized monitoring of animals without the need for tracking. This makes accelerometers especially useful for animals in outdoor production systems. Furthermore, the data collected by this device could be complemented with data available from other types of sensors, such as radio frequency identification tags, cameras, or microphones to improve the capabilities of the models.

**Conclusions**

This is the first study where the aim was to automate the detection of active and inactive states in rabbits. The results show that, with the device configuration, algorithms and hyperparameters tested, the models presented a high level of precision, signifying that the model is highly reliable in correctly classifying positive instances. Additionally, achieving a recall of 0.93 emphasizes the model's effectiveness in capturing a substantial portion of positive instances. The results could be improved if more data were gathered to have a better representation of the classes, which would also allow distinguishing between classes. Therefore, it appears to be a valuable tool for obtaining insights into the behaviour of rabbits and the factors that influence their welfare and productive efficiency.

**Acknowledgements**

We would like to thank Oscar Perucho for their contribution to data recording and animal care during the experiment. We would like to express our gratitude to Padraig Cunningham for his valuable grammar review contributions.

**Conflict of interest statement**

The authors declare that they have no known competing financial interests or personal relationships that could have appeared to influence the work reported in this paper.

## Author Contributions

Individual author contributions to this research article are as follows: Conceptualization and design: MM, MP, ID, methodology: MM, MP and LR, formal analysis: MM and LR, writing the original draft: MM, writing-review editing: MP, ID, LR, supervision: MP, LR and funding acquisition: MP, JPS.

## Data and model availability statement

The data that support the findings of this study are available from the corresponding author upon reasonable request.

# Supplementary Material

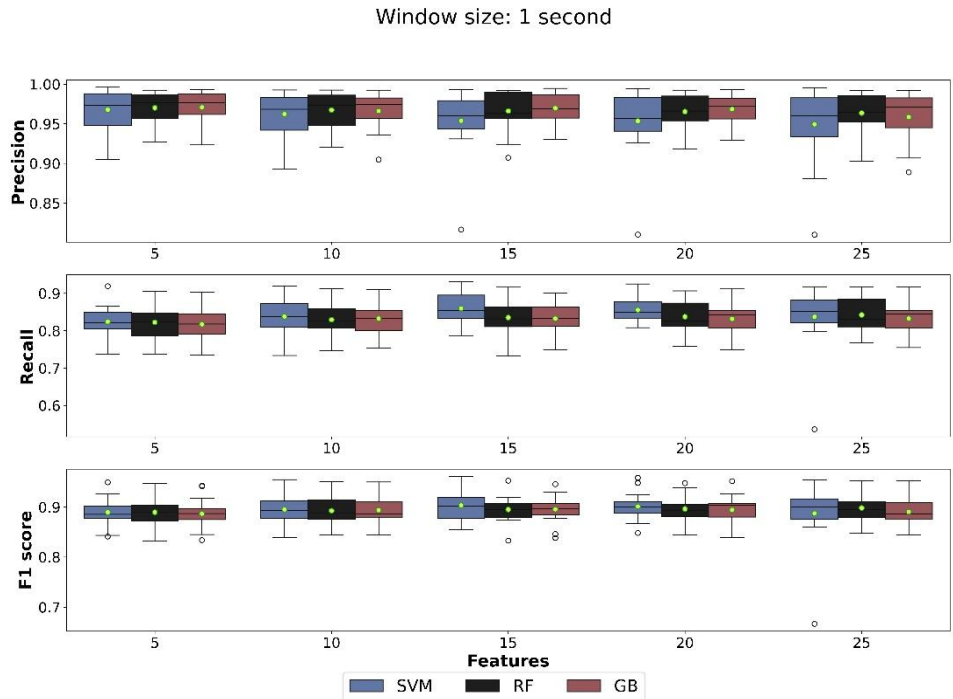

Figure S1. Distribution of precision, recall and F1 score for each model for each subset of the most important features with a window size of 1 second. The green dot represents the mean across the 16 outer loops.

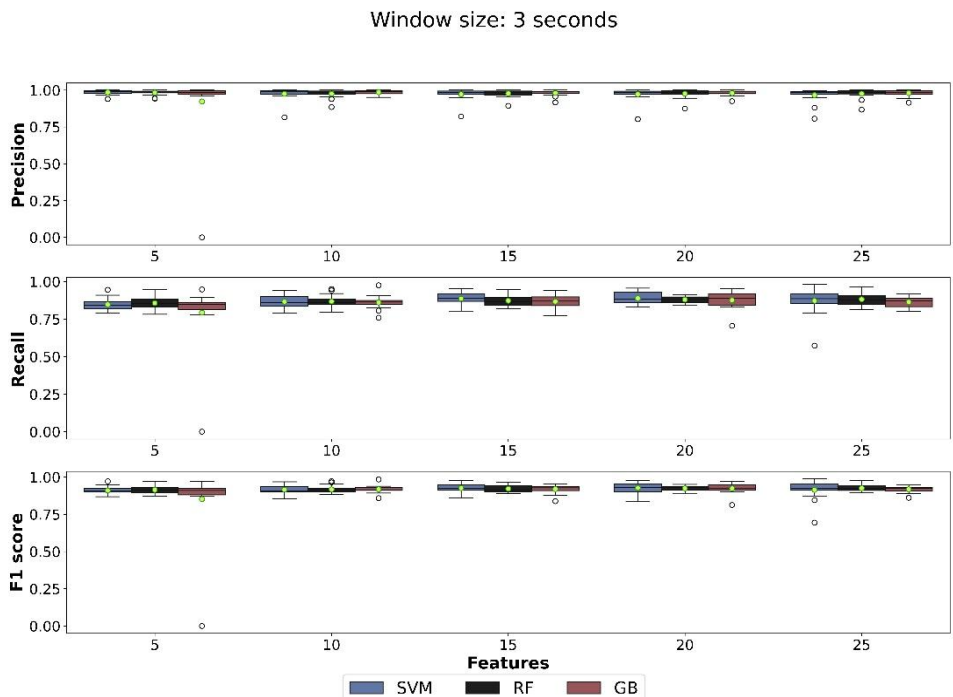

Figure S2. Distribution of precision, recall and F1 score for each model for each subset of the most important features with a window size of 3 seconds. The green dot represents the mean across the 16 outer loops.

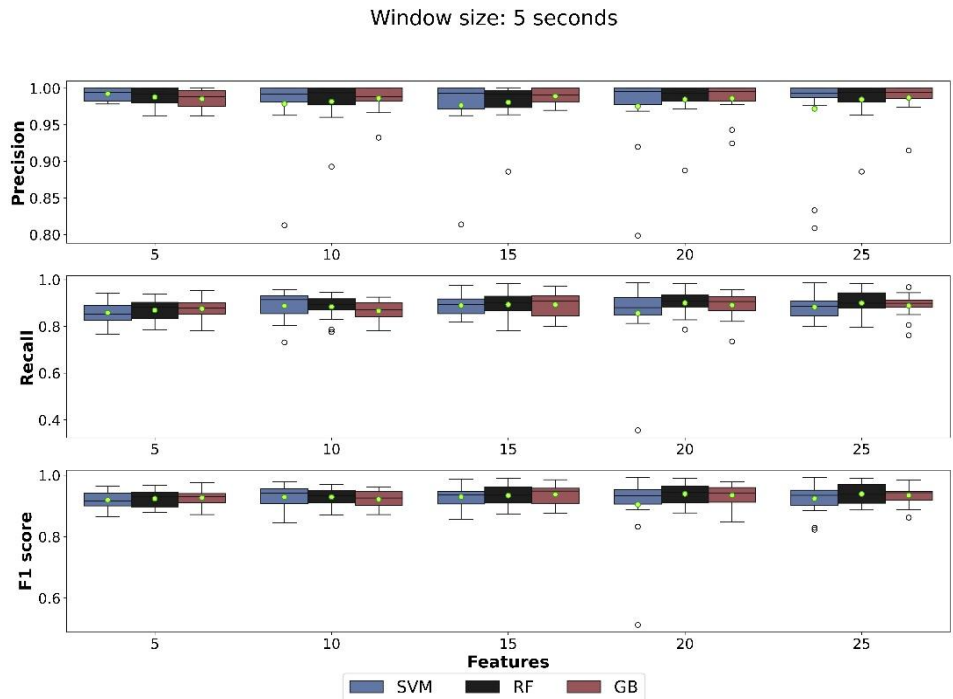

Figure S3. Distribution of precision, recall and F1 score for each model for each subset of the most important features with a window size of 5 seconds. The green dot represents the mean across the 16 outer loops.

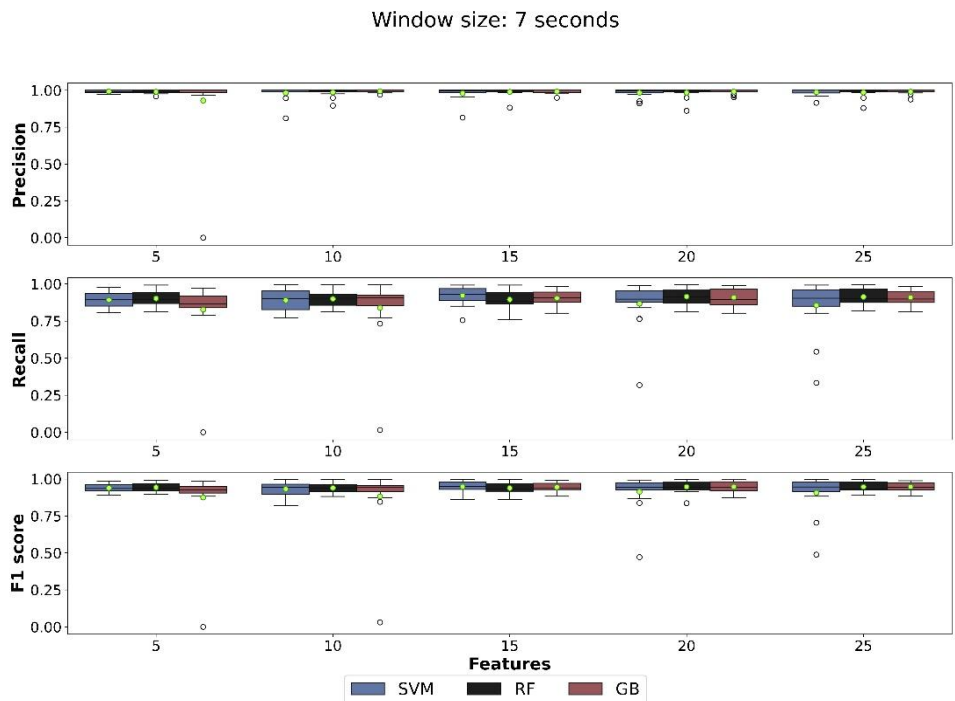

Figure S4. Distribution of precision, recall and F1 score for each model for each subset of the most important features with a window size of 7 seconds. The green dot represents the mean across the 16 outer loops.

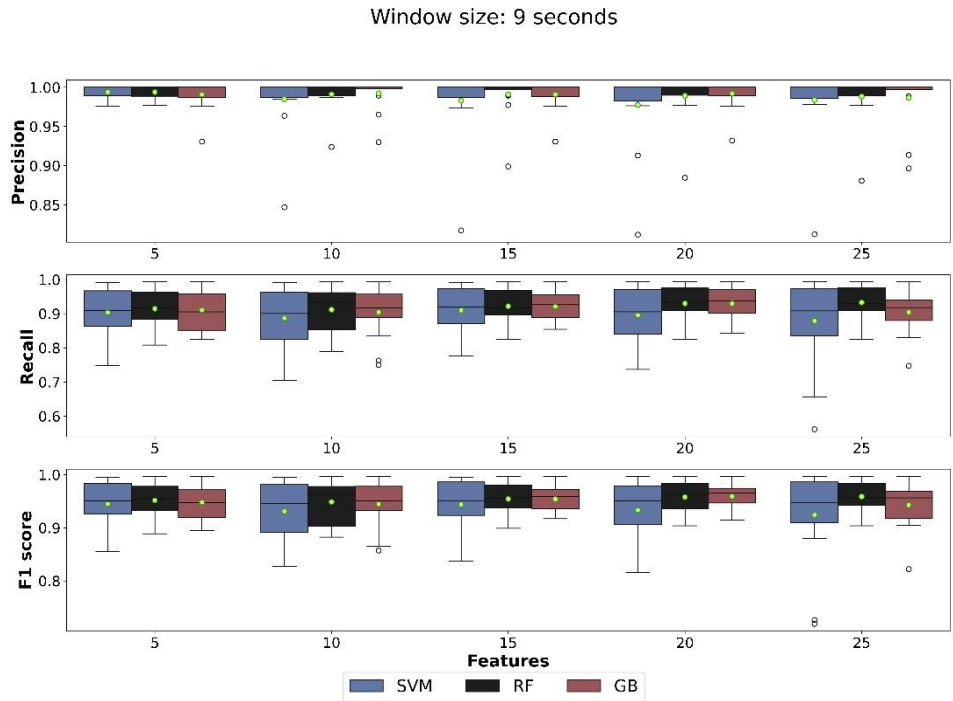

Figure S5. Distribution of precision, recall and F1 score for each model for each subset of the most important features with a window size of 9 seconds. The green dot represents the mean across the 16 outer loops.